\documentstyle[preprint,aps]{revtex}
\newcommand{\lra}{\longrightarrow}
\newcommand{\ce}{$\chi - \chi(Eu^{3+})$ }
\newcommand{\Tl}{$\rm T_{LT}$ }
\newcommand{\be}{\begin{eqnarray}}
\newcommand{\ee}{\end{eqnarray}}

\newcommand{\C}{$\rm Cu^{2+}$}

\newcommand{\Eu}{$\rm Eu^{3+}$ }
\newcommand{\Gd}{$\rm Gd^{3+}$ }

\newcommand{\Cc}{$ \chi_{spin}^{Cu^{2+}}$ }

\begin{document}
%*******************************************************
%
\title{Magnetism of the LTT phase of Eu doped $\bf La_{2-x}Sr_xCuO_4$}

\author{M. H\"ucker, J. Pommer,and B. B\"uchner}
\address{II. Physikalisches Institut, Universit\"at zu K\"oln,
50937 K\"oln, Germany}

\author{V. Kataev, and B. Rameev}
\address{Kazan Inst. for Technical Physics, RAS-420029 Kazan}

\date{\today}
\maketitle
\begin{abstract}
The ESR signal of Gd spin probes (0.5 at \%) as well as the static
normal state susceptibility of Eu (J(Eu$\rm ^{3+})=0$) doped
$\rm La_{2-x-y}Sr_xEu_yCuO_4$ reveal pronounced changes of the
Cu magnetism at the structural transition from the orthorhombic
to the low temperature tetragonal phase for all non--superconducting
compositions. Both a jumplike decrease of $\chi$ as well as the
ESR data show an increase of the in--plane magnetic correlation
length in the LTT phase. From the \Gd ESR linewidth we find that
for specific Eu and Sr concentrations in the LTT phase the
correlation length increases up to more than 100 lattice constants
and the fluctuation frequency of the $\rm CuO_2$ spin system slows
down to $\rm \sim 10^{10}- 10^{11}sec^{-1}$. However, there is no
static order above T$\sim$8K in contrast to the LTT phase of Nd
doped $\rm La_{2-x}Sr_xCuO_4$ with pinned stripe correlations.
\end{abstract}


\maketitle

%\begin{multicols}{2}

%\narrowtext

\section{Introduction}
The structural phase transition in rare earth doped
$\rm La_{2-x}Sr_xCuO_4$ (LSCO) from the orthorhombic (LTO)
to the low temperature (LT) tetragonal (LTT) phase has
attracted much attention during the last years,
since it manifests an intimate correlation between
the buckling pattern, superconductivity and magnetic
properties of the $\rm CuO_2$ planes in this compound
\cite{Buechner1,Buechner2,Axe}. In particular,
for Nd doped LSCO it is known that for compositions with
strongly suppressed superconductivity in the LTT phase
static antiferromagnetic order occurs \cite{Breuer2}.
Moreover, recent
neutron diffraction experiments by Tranquada et al.
\cite{Tranquada} give
evidence that for these non superconducting stoichiometries static
order at elevated temperatures
does occur in form of spatially modulated stripes of spins and holes.

In this paper we show that also in the non superconducting
LTT phase of Eu doped LSCO magnetism is affected dramatically
due to the LT transition. However, even at lowest temperatures
($\sim$8K) we do not find static order but extremely
slow antiferromagnetic dynamics. To our opinion the most
obvious reason for this different behavior is
a strong magnetic rare earth--Cu exchange interaction which is
absent in the case of the nonmagnetic \Eu ions. If this
explanation is correct, static order at elevated temperature
may not be intrinsic for the non superconducting LTT phase.

\section{Experimental}
Polycrystalline samples of $\rm La_{2-x-y}Sr_xEu_yCuO_4$ with
various Sr and Eu concentrations were prepared by a standard
solid state reaction described elsewhere \cite{Breuer}. The
role of Eu is
to induce the structural transition LTO $\lra$ LTT, whereas the
hole concentration is controlled independently by Sr doping.
The static susceptibility was measured for $\rm La_{1.8-x}Sr_x
Eu_{0.2}CuO_4$ with $0.04 \leq x \leq 0.17$ using a conventional
Faraday balance in a magnetic field of 1 Tesla.
The ESR experiments were performed at a frequency of 9.3GHz on $\rm
La_{1.99-x-y} Sr_xEu_yGd_{0.01}CuO_4$ samples using Gd$^{3+}$ ions
as spin probes. Such a small Gd concentration does not affect the
relevant physical properties as rare earth substitution is used in
any case to induce the LT transition.

\section{Results}
A representative curve of the temperature dependence of
the static magnetic susceptibility $\rm \chi(T)$ of
$\rm La_2CuO_4$
doped with Sr and Eu is shown in Fig.1 (upper curve). For the
analysis of $\rm \chi(T)$ it is necessary to account for several
contributions to the measured susceptibility.
In principal it consists of temperature
independent terms (such as core diamagnetism, Van Vleck
paramagnetism of $\rm Cu^{2+}$ ions), the spin $\rm S=\frac{1}{2}$ magnetism
of $\rm Cu^{2+}$ and the Van Vleck paramagnetism of $\rm
Eu^{3+}$:
\be
\chi(T)=\chi_0 + \chi^{Cu^{2+}}_{spin}(T) + \chi_{VV}^{Eu^{3+}}(T)
\ee
As we are interested in possible changes of the magnetism of the
CuO$_2$ planes as a function of temperature it is necessary to
subtract the dominant magnetic contribution of europium from the
total susceptibility curve. This can be done rather accurately
using a theoretical Van Vleck \cite{VanVleck} fit (see fig.1).
After its
subtraction the remaining signal $\rm \chi - \chi(Eu^{3+})$ shows a
jumplike decrease with lowering temperature at $\sim$123K which
clearly corresponds with the structural LT transition LTO $\lra$
LTT at $\rm T_{LT}$. The amplitude of this jump amounts to about
5\% of the
susceptibility of the corresponding Eu free $\rm La_{2-x}
Sr_xCuO_4$ compounds ($\rm \sim 1 \cdot 10^{-4} emu/mol$)
\cite{Takagi1}.
Furthermore, one can see a sign change of the temperature slope of the signal
immediately below $\rm T_{LT}$. It is known that the positive temperature
slope of \Cc in the LTO phase reflects the strong antiferromagnetic
Cu spin correlations in a two dimensional (2D) $\rm S = \frac{1}
{2}$ Heisenberg antiferromagnet. Hence, the observed drop of \Cc at
\Tl can be  naturally explained as being due to the larger magnetic
inplane correlation length $\rm \xi_{2D}$ in the LTT phase and
consequently to a smaller value of $ \chi_{spin}^{Cu^{2+}}$.

The Sr dependence of this anomaly in the susceptibility at the
structural transition is summarized in Fig.2. In this plot we show
changes of $\rm \chi - \chi (Eu^{3+})$ relative to a linear fit
of the data points in the LTO phase immediately above $\rm T_{LT}$.
With increasing
hole concentration the jump at \Tl becomes smaller and is scarcely
resolvable for x=0.17. From this we have to conclude that the
enhancement of the inplane correlation length $\rm \xi_{2D}$ due to the
structural changes at \Tl becomes smaller when the hole
concentration is increased. However, the absence of the jump in \Cc
for x$>$0.17 is not related to some qualitative change of the
structural LT transition, which is observable up to much higher Sr
concentrations (x$\sim$0.24). In contrast, we found that its
disappearance rather correlates with an electronic phase boundary
with\-in the LTT phase caused by a {\it critical buckling} of the
$\rm CuO_2$
planes \cite{Buechner1}. This phase boundary is related to a critical
Sr content
$\rm x_c$ that separates the LTT phase into a non superconducting
part ($\rm x<x_c$) and a superconducting part ($\rm x>x_c$).
Indeed, all these compounds presented in Fig.2 do not exhibit bulk
superconductivity, whereas without europium they do (except
x=0.04). It is worth to mention that the sample with x=0.17
already is very close to this critical concentration $\rm x_c$
as has been shown for a similar set of samples \cite{Buechner2}.

To summarize, the presented results of the static susceptibility
measurements
show that the LT transition significantly enhances the inplane
correlation length $\rm \xi_{2D}$, {\it but only if the LTT phase is not
superconducting}.

The same conclusion emerges from the ESR study.
The ESR spectra of $\rm La_{1.99-x-y}Sr_xEu_yGd_{0.01}CuO_4$ have
been measured in the temperature range 8K$<$T$<$300K. The fine
structure of the spectra (see inset in Fig.3) observed for all
samples is due to the small splitting of the ground state multiplet
$\rm ^8S_{7/2}$ of \Gd in the crystalline electrical field. The
width of the central component of the spectrum (encircled in inset
of Fig.3) as a function of temperature is shown in Fig.1 for two
representative $\rm La_{1.89-y}Sr_{0.1}Eu_yGd_{0.01} CuO_4$
compounds. One is not Eu doped and hence is a superconductor ($\rm
T_c\sim 25K$) without LTO $\lra$ LTT transition. For the second
sample with a very high Eu content (y=0.24) the LT transition
does occur at $\rm T_{LT}$$\sim$130K and superconductivity is
completely supressed. In the normal state of the LTO phase
$\rm\Delta H$ increases linearly with temperature for both samples.
This is due to the Korringa relaxation of Gd spins caused by a
small exchange coupling of \Gd ions to the mobile holes in the $\rm
CuO_2$ planes \cite{Kataev1}. The important feature of
the Eu doped sample is a very strong increase of the linewidth at
low temperatures, i.e. in the LTT phase.

To explain such a profound effect it is necessary to consider a
coupling of the Gd spins to antiferromagnetic (AF) spin fluctuations
in the $\rm CuO_2$ planes. Due to the magnetic exchange interaction
between \Gd and \C ions ($\rm J_{Cu-Gd}$) the relaxation rate $\rm
1/T_1$ of the \Gd spin state serves as a probe for the Cu spin
fluctuation frequency $\rm \omega_{sf}$: $\rm 1/T_1\propto
J_{Cu-Gd}^2 \chi_{Cu}T/\omega_{sf}$. Since $\rm 1/T_1$ determines
the homogeneous part of the width $\rm \Delta H$ of the components
of the Gd ESR spectrum, estimates of $\omega_{sf}$ can be obtained
from the temperature dependence of $\rm \Delta H$ \cite{Kataev2}.
In the LTO phase AF fluctuations with the frequency of the order of
$\rm 10^{13}\, sec^{-1}$ do not contribute significantly to the Gd
ESR linewidth. Thus, a strong increase of $\rm \Delta H$
corresponds to a dramatic slowing down of the \C spin dynamics in
the LTT phase. The estimated fluctuation frequency decreases by two
orders of magnitude down to $\rm \omega_{sf}\sim
10^{10}-10^{11}sec^{-1}$ at T$\sim$10K \cite{Kataev2}. If we assume
that $\rm \omega_{sf}\sim 1/\xi_{2D}^2$ \cite{Millis}, then
the magnetic correlation length $\rm \xi_{2D}$ of the hole doped
$\rm CuO_2$ planes should increase up to more than 100 lattice
constants. Remarkably, the continuous increase of the linewidth
$\rm \Delta H$ down to the lowest temperatures of the measurements,
indicates that the Cu spins remain dynamic rather then turning into
an AF ordered state.

\section{conclusion}

In conclusion, we have shown that in the LTT phase of Eu doped
$\rm La_{2-x}Sr_x CuO_4$ both the static susceptibility and the \Gd
ESR linewidth indicate a pronounced increase of the Cu spin
correlation length for all {\it non} superconducting samples.
However, in contrast to the observation of static stripe order of
spins and charges in the LTT phase of Nd doped $\rm La_{2-x}Sr_x
CuO_4$ below 50K, for the Eu doped compounds we find no evidence
for a static order even at 8K.

This work was supported by the Deutsche Forschungsgemeinschaft
through SFB 341. M.H. acknowledges support by the
Graduiertenstipendium des Landes Nordrhein--Westfalen.
The work of V.K. and B.R. was supported in part by
the State HTSC Program of the Russian Ministry of Sciences (project
No.940045) and by the Russian Foundation for Basic Research
(project No.95-02-05942). V.K. acknowledges NATO support under
Collaborative Research Grant No. HTECH.EV 960286.

\newpage
%Fig. 1
\begin{figure}
%\begin{center}
%\pictspace{85}{102}
%\pict{85}{102}{0}{112}{fig1.pcx}
\caption[]{Static susceptibility of $\rm La_{1.76}Sr_{0.04}Eu_{0.2}CuO_4$.
           The total measured signal $\chi$, the Van
           Vleck fit $\chi(Eu^{3+})$ which takes the Eu magnetism into
           account and the difference \ce are shown. Inset: changes
           of \ce at the LTO $\lra$ LTT transition on expanded scale.}
%\end{center}
\end{figure}
%
%Fig. 2
\begin{figure}
%\begin{center}
%\pictspace{85}{100}
%\pict{85}{100}{-4}{110}{fig2.pcx}
\caption[]{Deviations of the static magnetic susceptibility $\rm
            \chi-\chi(Eu^{3+})$ of $\rm La_{1.76}Sr_{0.04}
            Eu_{0.2}CuO_4$ with 0.04$<$x$<$0.17 from a linear fit
            in the LTO phase (curves are shifted). The
            jumplike anomly at $\rm T_{LT}
            \sim 130K$ is due to the LT transition and decreases
            with decreasing buckling of the $\rm CuO_2$ planes.
            All samples are no bulk superconductors.}
%\end{center}
\end{figure}
%Fig. 3
\begin{figure}
%\begin{center}
%\pictspace{85}{105}
%\pict{85}{105}{-4}{115}{fig3.pcx}
\caption[]{Temperature dependence of the linewidth $\rm \Delta H$
           of the encircled component (inset) of the $\rm Gd^{3+}$
           ESR spectrum of $\rm La_{1.89-y}Sr_{0.1}Eu_yGd_{0.01}CuO_4$
           with y=0.0 and 0.24. Remarkable is the pronounced broadening
           of the spectra in the LTT phase of the non superconducting
           sample. For this sample the contribution due to thermally
           excited states of \Eu laying above 400K has been subtracted.}
%\end{center}
\end{figure}
%\end{multicols}
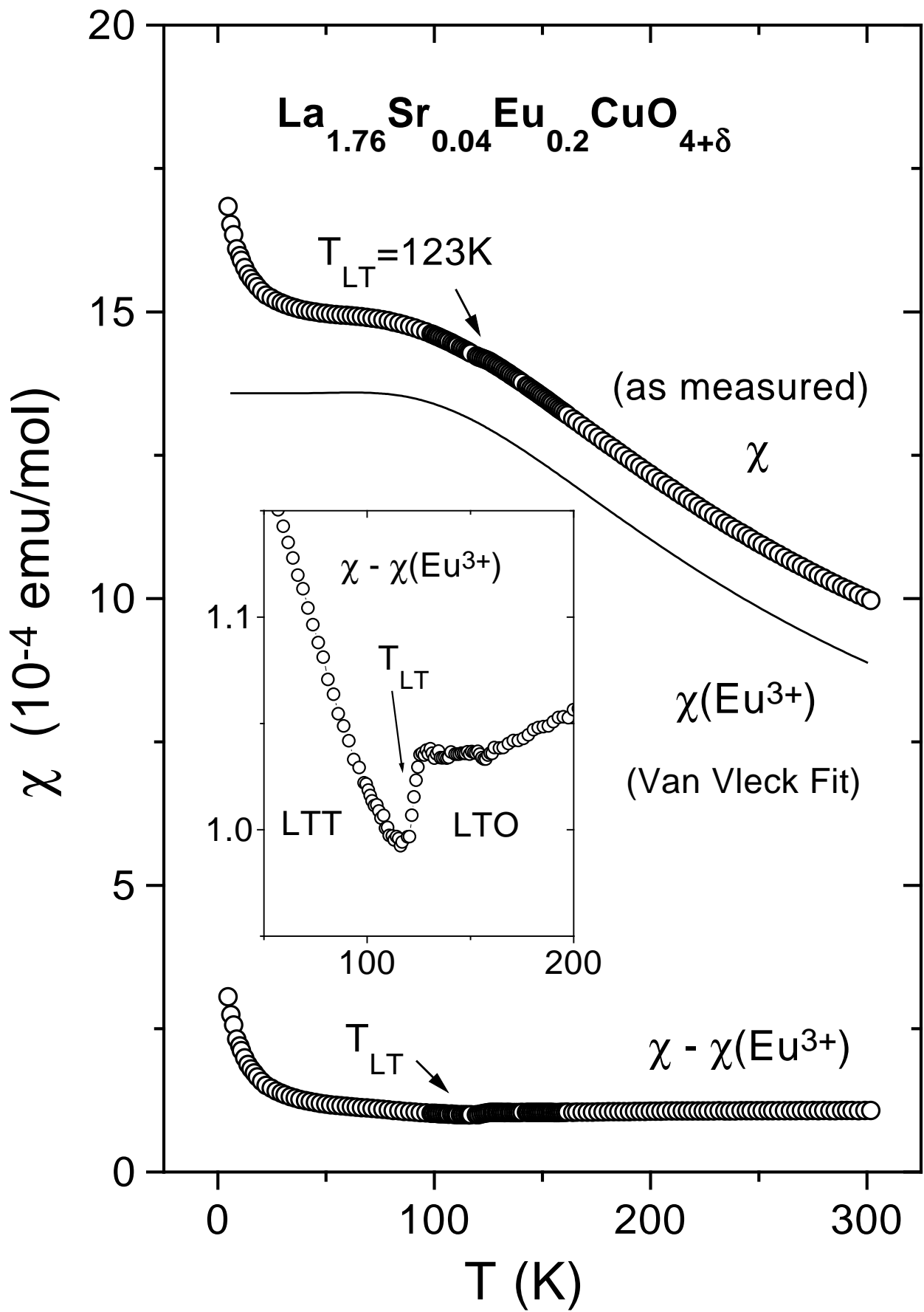
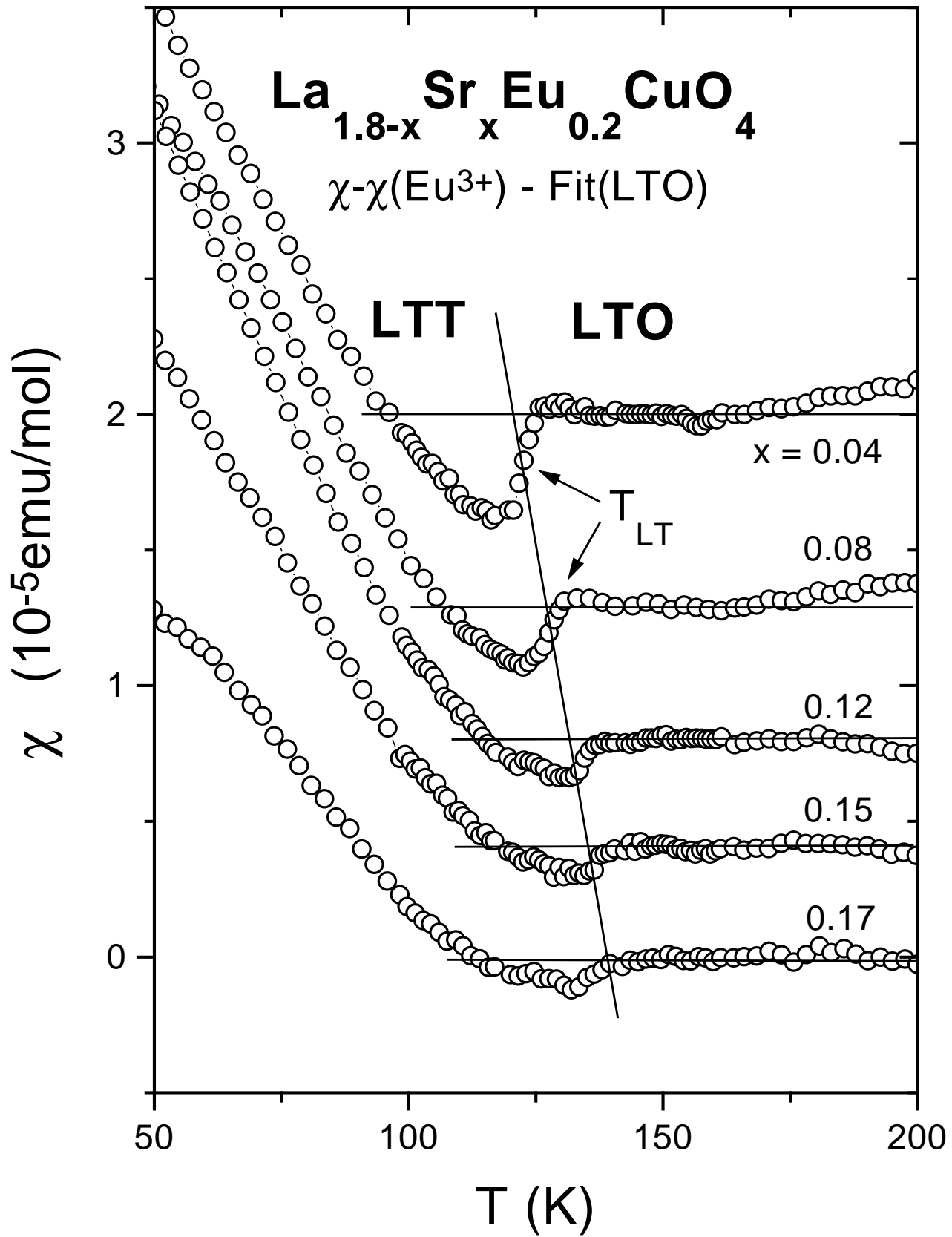
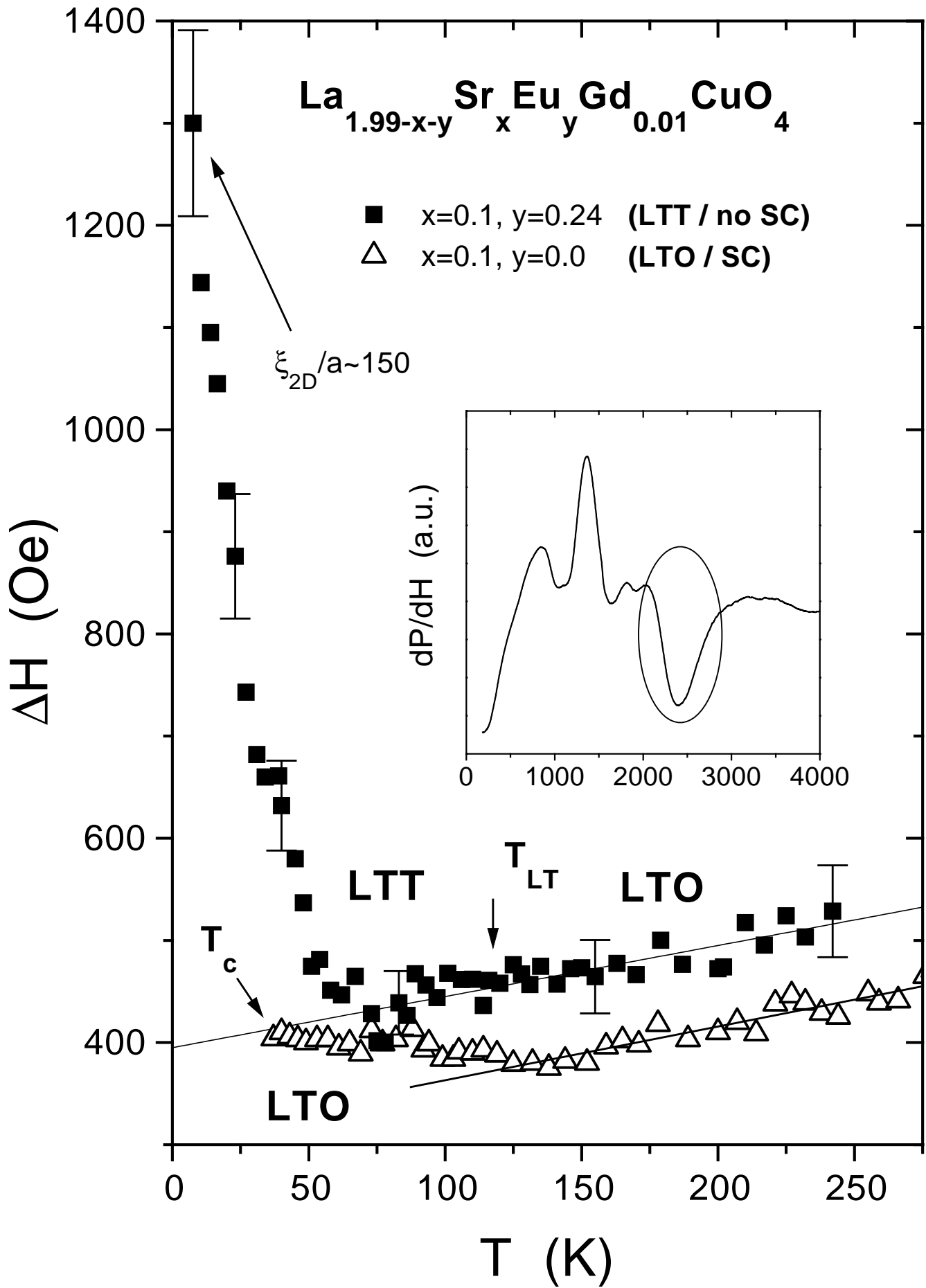
\end{document}